# Breaking the Alphabet: Rethinking File Ordering in Code Review

Md Shamimur Rahman
Computer Science
University of Saskatchewan, Canada
Saskatoon, Saskatchewan, Canada
mdr614@usask.ca

Zadia Codabux
Computer Science
University of Saskatchewan, Canada
Saskatoon, Saskatchewan, Canada
zadiacodabux@ieee.org

Chanchal K. Roy
Computer Science
University of Saskatchewan, Canada
Saskatoon, Saskatchewan, Canada
chanchal.roy@usask.ca

## Abstract

Effective code review is central to maintaining software quality, yet there is limited research about how the ordering of changed files in Pull Requests (PRs) influences review effectiveness. Most popular code review tools default to alphabetical ordering, favoring predictability over contextual relevance. While prior studies examined how file position shapes reviewer attention, it remains unclear how such ordering influences cognitive load and perceived review thoroughness. This study presents the first large-scale survey of 1,355 professional developers across 182 widely used open-source projects to investigate how file ordering impacts review behavior, comprehension, and perceived effectiveness. Our mixed-methods analysis reveals that only 10.2% of reviewers consider alphabetical ordering optimal, underscoring a cognitive misalignment in their interpretation of code changes. Although some developers appreciate its predictability, more than half (57.6%) report that it increases context switching, disrupts logical reasoning, and contributes to review fatigue, and 63.9% expressed concern that the default ordering may cause them to miss bugs. We further identify key challenges in multi-file reviews and elicit developers' expectations for improved tooling, including dependency-aware grouping and customizable file ordering (requested by 66% of reviewers). These findings highlight the need for reviewer-centric interface designs that better align tool behavior with human cognition.

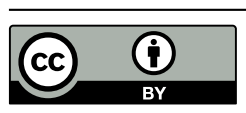

## 1 Introduction

Peer code review is a widely adopted practice in software development, where code authored by a developer is examined by peers before integration into the main codebase. This collaborative process plays a critical role in maintaining software quality by enabling early detection of defects, technical debt, and code smells, as well as team-wide knowledge sharing and adherence to coding standards and project-specific guidelines [9, 13, 17, 18, 61, 64, 85, 91]. Leading tech companies such as Microsoft [5], Google [76], and Oracle [24], as well as prominent open-source projects like Android, Qt, OpenStack [65], and Eclipse [60], have made code review a core part of their development workflows. While early code reviews were conducted through in-person or checklist-driven inspections [33], Modern Code Review (MCR) practices leverage collaborative tools [15, 16, 65, 87] that support asynchronous, distributed collaboration across geographically dispersed teams [5]. In MCR, developers submit Pull Requests (PRs), logical sets of changes that introduce new features, enhancements, or bug fixes, which are then evaluated by experienced developers or maintainers. Popular tools like GitHub [15] and Gerrit [65] typically present modified files in alphabetical order by file path. For example, a PR modifying *src/auth/login.js*, *core/utils.js*, and *src/ui/button.js* would display them in the order: *core/utils.js*, *src/auth/login.js*, *src/ui/button.js*.

Prior studies show that the position of changed files during code review significantly influences reviewer behavior, with earlier files receiving more attention and comments, a phenomenon known as positional bias [7, 38]. Bouraffa et al. [19] further observed that alphabetical ordering, the default in most tools, often misaligned with reviewers' natural inspection preferences. In their study of over 23K PRs, 44.6% exhibited commenting patterns that deviated from the alphabetical sequence. Other works proposed alternative ordering strategies, including prioritizing files with larger code changes, salient classes, test files, or those semantically linked to the PR's title and description [7, 19, 48]. However, the effectiveness of these strategies varies across review contexts and individual preferences. What remains essential is that the file order neither compromises review quality nor leads to overlooked bugs. This concern is grounded in cognitive factors tied to file positioning. Prior works highlight issues such as attention decrement, i.e., diminishing focus across sequential items [88], and working memory limitations, which affect the ability to retain and process information during complex tasks [10, 62, 82, 94]. These effects are especially pronounced in lengthy reviews, where files later in the sequence may receive suboptimal scrutiny due to mental fatigue.

Given this cognitive framing and growing empirical evidence of positional bias, it is crucial to examine whether the order of files, particularly default alphabetical ordering, increases reviewers' cognitive burden and consequently affects their likelihood of missing defects or reduces overall review efficiency. Additionally, there is limited understanding of the usability challenges developers face when navigating multi-file PRs in practice, including whether they follow default file ordering, adopt alternative navigation strategies, and how these behaviors influence their review experience.

Fregnan et al. [38] conducted a controlled experiment with 106 developers, introducing unrelated defects into files placed first and last in a review sequence. Developers were 64% less likely to detect defects in the last-positioned file and spent significantly less time reviewing it, demonstrating positional bias in defect detection. However, this controlled setting does not reflect the complexity of real-world reviews, which involve multi-file PRs, time pressure, and contextual factors [32, 45, 96].

To bridge this gap, we conducted a large-scale survey of 1,355 professional developers representing 67 countries and 182 popular open-source projects. The survey investigates how file ordering influences review behavior, perceived effectiveness, cognitive effort, and overall usability in multi-file reviews. In particular, we examine whether the default alphabetical ordering used in most review tools aligns with developers' cognitive strategies and preferences. Participants reported their file-prioritization practices, perceptions of alphabetical ordering, and the challenges they encounter during multi-file reviews. Our findings aim to inform the design of review tools that better accommodate human cognition and support more effective review practices. Overall, our contributions are as follows:

- **Empirical characterization of file-ordering practices:** We provide large-scale empirical evidence from 1,355 developers across 67 countries and 182 open-source projects that reveal how reviewers navigate multi-file PRs and the extent to which they adhere to or deviate from default alphabetical ordering.
- **Perception and usability analysis of ordering preferences:** Our quantitative and qualitative analyses show how reviewers perceive alphabetical ordering across efficiency, accuracy, and cognitive fit, and uncover key factors shaping these perceptions.
- **Comprehensive taxonomy of multi-file review challenges:** We identify and organize both general review challenges and ordering-specific issues, showing how these difficulties vary across experience levels and contribute to cognitive load and review inefficiency.
- **Identification of expectations for improved tooling:** We derive actionable implications for reviewer-centric tool design, emphasizing usability through support for dependency-aware grouping and cognitively aligned file prioritization.
- **Open replication package:** We release the survey instrument, analysis scripts, and supplementary materials in an open replication package[1] to support reproducibility and future research.

## 2 Background and Related Work

This section provides an overview of the contextual background and reviews relevant literature across the following key areas.

### 2.1 Cognitive Constraints in Code Review

While reviewers support software extensibility and defect mitigation, the effectiveness of code review is limited by cognitive constraints and biases that hinder the accurate evaluation of code changes [5, 76]. Cognitive load theory suggests that review tasks impose substantial mental demands, including sustained attention, mental model construction, reasoning about delocalized bugs, and comprehension of large or multi-file changes, often exceeding working memory capacity [11, 12, 41, 42]. Beyond general cognitive load, other cognitive constraints, including confirmation bias, anchoring effects, availability bias, and representativeness heuristics, could deviate reviewers from rational judgment processes and compromise review effectiveness [22, 49, 82, 86]. For instance, Huang et al. [49] conducted an eye-tracking and medical imaging study to explore how biases manifest during the inspection process, revealing that developers often rely on familiar patterns or recent experiences when assessing unfamiliar code. Moreover, attention switching, decision fatigue (e.g., inattention, impulsivity, procrastination), and a mismatch between reviewer expertise and code further lead to slower reviews and degrade review effectiveness [42, 50].

### 2.2 Psychology of File Position in PR Review

Human cognitive and psychological limitations influence the thoroughness of code reviews, particularly in how attention is distributed across files within a PR. Baum et al. [11] found that reviewers tend to begin at the top of the file list, leading to uneven attention. In a large-scale study, Fregnan et al. [38] demonstrated that top-positioned files receive more comments. In a controlled experiment involving 106 participants, bugs in the first file were 64% more likely to be detected than those in the last. Bagirov et al. [7] further confirmed that reviewers focus disproportionately on earlier files in closed-source industrial projects. These findings highlight how attention diminishes over the review sequence, a phenomenon called attention decrement, where sustained focus declines over time, particularly with sequential tasks [46, 66, 88]. This decay introduces positional bias, challenging the assumption that all files receive equal scrutiny. Such uneven attention distribution can influence review effectiveness, for instance, McIntosh et al. [61] showed that differences in review coverage correlate with software quality, highlighting the practical consequences of reviewer attention imbalance. Another key psychological factor influencing review performance is the limited capacity of human working memory, which involves processing and manipulating information during complex tasks [6, 12, 94]. While individual capacity varies, research consistently shows that working memory can only maintain a limited amount of information [25, 26]. This limitation plays a key role in cognitively demanding programming tasks, such as understanding deeply nested logic or tracking non-local dependencies [14]. Moreover, in multi-file PRs, as cognitive load accumulates, reviewers may struggle to retain context across files, reducing attention and comprehension as the review progresses.

### 2.3 Investigation on Reviewing Code Changes

To decide whether to accept, reject, or request changes to a PR, reviewers must understand code modifications in context and assess their necessity and quality. Baum et al. [11] found that reviewers often follow the default file order, typically alphabetical by file path, though later work suggests more effective strategies. Gonccalves et al. [42] recommend grouping related changes, while Bagirov et al. [7] showed that ordering files by change size surfaces critical or error-prone components earlier than alphabetical sorting.. Bouraffa et al. [19] found that 44.6% of reviewers deviated from alphabetical order, often starting with large diffs, files tied to the PR description, or tests when both test and production code are present. Olewicki et al. [70] proposed a similarity-based ordering to highlight files needing attention, though navigation patterns varied with reviewer familiarity and preferences. Beyond file order, eye-tracking research reveals that certain code elements inherently attract reviewer focus independent of their location. Abid et al. [1] found that developers primarily focus on function calls, followed by control flow constructs and method declarations, consistent with Rodeghero et al. [75], who identified method signatures as the primary focal point for reviewers, with substantial attention also given to call terms and control structures. Additionally, Al Madi et al. [2] showed that tokens with low frequency and higher character length tend to hold reviewers' attention longer, suggesting that both semantic relevance and lexical complexity influence visual engagement.

[1] https://doi.org/10.5281/zenodo.16111982

Prior work shows that file position affects reviewer attention and navigation behavior in multi-file PRs, including controlled studies on comment distribution and defect detection [38] and analyses of non-alphabetical navigation patterns in reviewer commenting sequences [19]. However, existing studies primarily provide observable behavioral evidence from controlled experiments or repository analyses, without explaining the reviewers' perceptions, reasoning, and subjective experiences underlying reviewers' navigation behaviors. This study addresses this gap through a large-scale survey that captures practitioners' perceptions of default alphabetical ordering, coping strategies, and usability challenges in multi-file reviews, as well as their expectations for improved tool support.

# 3 Methodology

Building on the identified gaps, this section outlines the research questions and describes the survey used in this study.

## 3.1 Research Questions

To address the research gap on file ordering, this study conducts a large-scale survey of active open-source contributors and professional developers. Our investigation is guided by the following Research Questions (RQs):

**$RQ_1$. *How do reviewers navigate multi-file PRs during code review?*** Understanding reviewers' navigation behavior characterizes their interaction with review interfaces. While most platforms default to alphabetical ordering, reviewers may adopt alternative sequences. This RQ examines the extent to which reviewers follow or deviate from the default file ordering and how their navigation practices align with tool-imposed ordering.

**$RQ_2$. *How do reviewers perceive default alphabetical file ordering?*** Perceptions of ordering optimality reflect reviewers' judgments about efficiency, accuracy, and cognitive fit. Since cognitive factors such as attention decrement [46, 66, 88] and working memory limitations [6, 12, 94] affect review thoroughness, understanding how developers perceive and evaluate alphabetical ordering is essential. This RQ further explores the factors shaping these perceptions and their implications for designing cognitively supportive ordering strategies. • **$RQ_{2.1}$** *What are reviewers' perceptions of alphabetical ordering optimality?* • **$RQ_{2.2}$** *What underlying factors influence these perceptions?* • **$RQ_{2.3}$** *Why do reviewers find alphabetical ordering suboptimal or acceptable?*

**$RQ_3$. *How is file ordering associated with reviewers' perceived review effectiveness and quality?*** Misalignment between file ordering and reviewers' cognitive preferences may prompt alternative reordering, potentially shifting effort, attention, and accuracy during review. This RQ investigates whether and how such misalignment imposes burdens and undermines review quality.

**$RQ_4$. *What challenges do reviewers face when reviewing multi-file PRs?*** While reviewing small PRs (1–3 changed files) may not substantially increase cognitive load, larger multi-file changes can strain reviewers' attention, context retention, and reasoning across inter-file dependencies. This RQ investigates which challenges emerge as PRs span multiple files and how these difficulties relate to file ordering. • **$RQ_{4.1}$** *What general challenges do reviewers encounter in multi-file reviews?* • **$RQ_{4.2}$** *What difficulties do reviewers attribute to or experience in relation to suboptimal file ordering?*

**$RQ_5$. *What are reviewers' expectations for improving multi-file PR reviews?*** Building on identified cognitive and usability challenges, this RQ explores what improvements reviewers consider most meaningful, particularly regarding customizable file ordering, to inform design requirements for the next-generation review tools.

## 3.2 Survey Design and Data Analysis Plan

To address the RQs, we conducted an online survey following established empirical software engineering guidelines [73], as described in the following subsections.

*3.2.1 Survey Design.* The survey questionnaires were designed to ensure clarity, relevance, and consistency with the study objectives. It includes both closed- and open-ended questions covering key aspects of multi-file code review. Topics include review strategies, perceptions of default file ordering, challenges in reviewing multi-file PRs (e.g., cognitive load, change impact), adaptive behaviors (e.g., mental reordering, external tool use), and preferences for alternative ordering mechanisms. We also collected basic demographic and professional information, including review experience, review frequency, and tools used. The survey materials are available in our replication package[1].

*3.2.2 Survey Validation and Participant Recruitment.* To ensure face validity[2], the survey was reviewed by two industry practitioners (with 8 and 17 years of experience) and two software engineering researchers (with 20+ years in software engineering and 6+ years in code review research). Iterative revisions followed, and the survey received approval from the University of Saskatchewan Ethics Behavioral Board.

Recruiting software developers with code review experience for online surveys is challenging. We targeted contributors from active open-source projects with substantial review activity, selecting 182 mature repositories from the Apache Software Foundation and GitHub using the SEART[3] search engine based on popularity and activity metrics (stars, contributors, commits, and recent PRs). From these projects, we collected public email information for over 7,000 eligible developers and successfully delivered the survey to 6,157 valid addresses. To increase participation, we also used snowball sampling, encouraging respondents to share the survey within their professional networks. Detailed project selection criteria and recruitment statistics are provided in the supplementary material[1].

[2]https://bit.ly/Validating_a_Survey
[3]https://seart-ghs.si.usi.ch/

*3.2.3 Analysis Plan.* We employed a mixed-methods approach tailored to survey question types. The quantitative analysis of closed-ended responses used descriptive statistics, chi-square tests with Cramér's V for effect sizes [27], and binary/ordinal logistic regression [47] controlling for demographic and experiential covariates. Model diagnostics included pseudo-$R^2$ and checks for potential multicollinearity among predictors [52, 80]. All statistical significance tests used $\alpha = 0.05$ with odds ratios and 95% confidence intervals reported [69, 84]. The qualitative analysis combined manual categorization and thematic analysis [20], both independently conducted by two evaluators through open coding. Concise factual responses (e.g., navigation strategies) were categorized into predefined or emergent groups and summarized by frequency, while narrative responses on perceptions and challenges were thematically coded to derive higher-level concepts. Inter-rater reliability (Cohen's $\kappa >$ 0.80) and consensus discussions ensured agreement, and qualitative insights were triangulated with quantitative findings [23].

# 4 Evaluation and Results

In this section, we present our findings and provide insights into reviewers' navigation behavior, cognitive challenges, and expectations in multi-file code reviews.

## 4.1 Overview of Participant Demographics

Over a two-month period, the survey received 1,392 responses. After excluding 14 submissions without consent and 23 entries that were substantially incomplete, we analyzed 1,355 valid responses, which formed the basis of this study. These responses are from participants representing 67 countries, providing broad geographic coverage. Europe contributed the most (49.4%, 669), followed by North America (29.7%, 402), Asia (13.4%, 181), Oceania (4.3%, 58), South America (2.1%, 28), and Africa (1.3%, 17). The top five countries were the United States (22.7%, 308), Germany (11.0%, 149), Canada (6.9%, 94), the Netherlands (4.4%, 60), and the United Kingdom (4.4%, 60). Together, the top 10 countries accounted for 66.2% of responses, and the top 20 for 87.0%, with the remaining 13.0% distributed across 47 other countries.

Participants had varying levels of professional experience: 48.9% (663) had more than 16 years of software development experience, 19.3% (262) had 12–15 years, and 17.0% (230) had 8–11 years. Only 3.1% (42) had less than 4 years of experience. Overall, 85.2% had more than 7 years of development experience. In terms of code review experience, 28.9% (391) had 8–11 years, 24.0% (325) had 4–7 years, 18.6% (252) had over 16 years, and 18.1% (245) had 12–15 years. Only 10.5% (142) had 0–3 years of review experience, indicating a highly experienced respondent pool. Regarding participation frequency, 58.0% (786) engaged in code reviews daily, 26.7% (362) weekly, and the remainder less frequently: bi-weekly (4.6%), monthly (4.4%), or rarely (6.3%). These insights show that code review is a regular practice for most respondents. Next, we asked participants to identify their preferred code review platforms. Most reported using cloud-based Git platforms, with GitHub dominating (70.4%, 869), followed by GitLab (13.9%, 172) and Bitbucket (5.1%, 63). Azure DevOps appeared in 30 open-text responses. Several respondents also mentioned internal tools used in large organizations, such as Critique (Google), CRUX (Amazon), and Phabricator (Meta). Others used specialized tools like Graphite, Gerrit, and Review Board, as well as IDE-based integrations (e.g., IntelliJ, VS-Code, GitLens) and traditional workflows (e.g., mailing lists, CLI), reflecting diverse review environments.

## 4.2 $RQ_1$- Multi-file PRs Navigation

To understand how reviewers navigate multi-file PRs, the survey presented common strategies (e.g., default alphabetical order, change type, change size) and allowed for custom responses. We analyzed 1,350 responses, including 382 open-ended comments. Two independent evaluators (each with over eight years of programming experience) manually coded the open responses, achieving high agreement ($\kappa = 0.93$) [81]. Disagreements were resolved by a third evaluator. Of the 382 comments, 377 were successfully coded, and five were excluded due to insufficient detail.

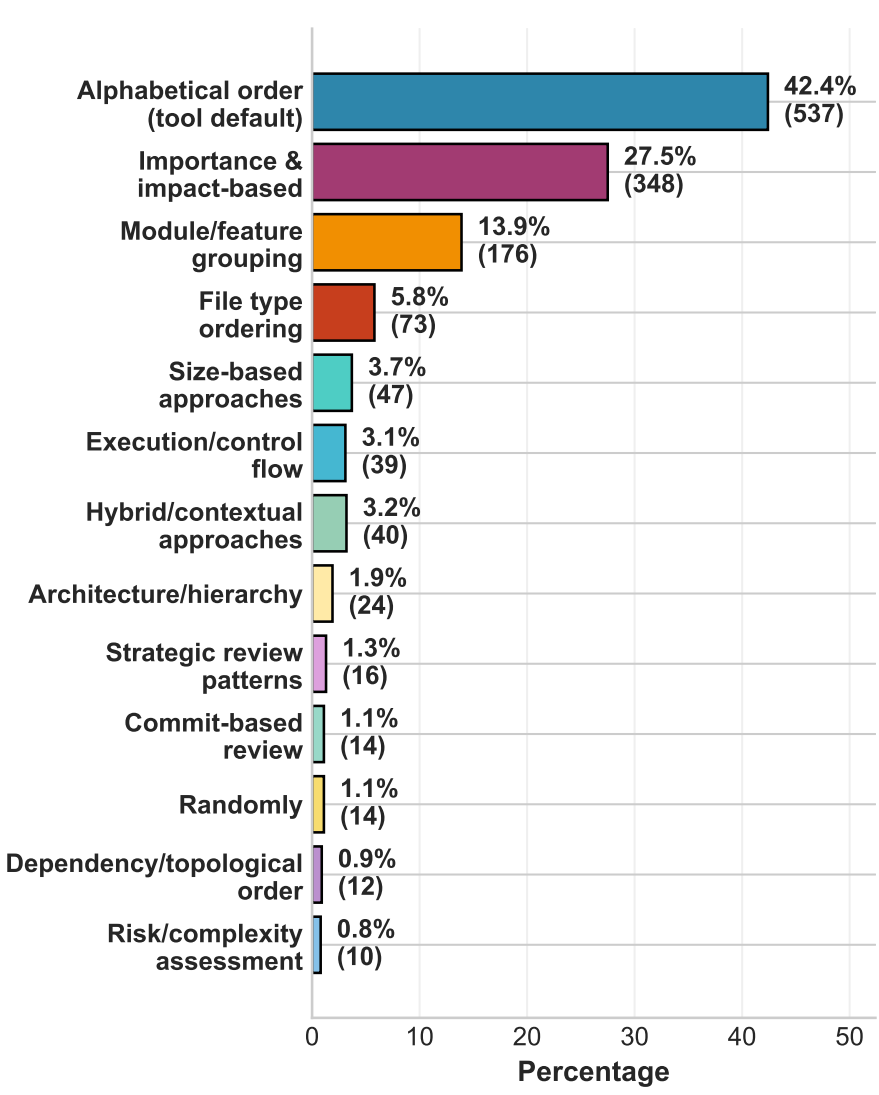

**Figure 1: Multi-File Review Strategies Mentioned in Survey**

Although review tools typically display changed files alphabetically, our tool-specific analysis shows that many reviewers consciously deviate from this ordering. Alphabetical sequencing remained the most frequent approach among users of GitHub (39.6%, 344), GitLab (43.0%, 74), Bitbucket (28.6%, 18), Azure DevOps (30.0%, 9), and other tools (41.6%, 92). Yet a substantial number of reviewers in each platform prioritized files by importance or system impact (e.g., reviewing core logic or security-related code before auxiliary components) on GitHub (30.0%, 261), GitLab (27.3%, 47), and Bitbucket (27.0%, 17), or grouped related files within the same module or feature, including GitHub (15.2%, 132) and Azure DevOps (17.2%, 5). Overall, we identified 13 distinct strategies (Figure 1), with smaller subsets ordering by file type (5.8%) such as programming, non-programming, test files, or reviewing smaller changes first (3.7%), or following control flow (3.1%). Less frequent approaches involved hybrid strategies and commit-based navigation. While alphabetical ordering remains most common (42.4%), most participants (57.6%) preferred alternative approaches. These variations indicate that, although alphabetical sequencing remains the prevailing default across review tools, many developers adopt impact-driven or context-based ordering strategies that better align with comprehension needs and project context.

**Answer for $\mathbf{RQ_1}$–** While 42.4% of reviewers adhere to the default alphabetical file ordering, evidence shows that the majority adopt alternative, context-driven strategies rather than the tool-imposed order when reviewing multi-file PRs.

## 4.3 $\mathbf{RQ_2}$- Perceptions of Alphabetical Ordering

Building on the navigation behaviors identified in $\mathbf{RQ_1}$, we investigate reviewers' perceptions of the default alphabetical file-ordering strategy. We combined descriptive analysis with logistic-regression modeling to examine how reviewers' experience relates to their perceptions of file ordering and behavioral intentions.

*4.3.1* $\mathbf{RQ_{2.1}}$ *- Perceptions of Optimality.* We asked participants to evaluate whether they consider the default alphabetical file ordering in code review tools to be optimal, i.e., effective and suitable for supporting review comprehension and navigation. Among the 362 respondents who reported using alphabetical ordering, only 9.9% considered it optimal, while 52.2% rated it as suboptimal and 37.8% were neutral ($\chi^2(2) = 61.87$, $p < .001$). A similar pattern emerged among the 707 reviewers who employed alternative navigation strategies: 10.3% (73) viewed alphabetical ordering as optimal, 54.7% (387) as suboptimal, and 34.9% (247) as neutral ($\chi^2(2) = 122.77$, $p < .0001$). When combined, 10.2% (109) of all respondents perceived alphabetical ordering as optimal, compared to 53.9% (576) who regarded it as suboptimal and 35.9% (384) who remained neutral ($\chi^2(2) = 184.64$, $p < .0001$, Cramér's V = 0.29, medium effect), indicating a moderate and statistically significant preference against alphabetical ordering. A subsequent chi-square test of independence revealed no significant difference in perceptions between those who use alphabetical ordering and those who adopt alternative strategies ($\chi^2(2) = 0.37$, $p > .05$), suggesting that dissatisfaction with alphabetical ordering is pervasive regardless of whether reviewers actively avoid it or continue using it by default.

> **Finding 2.1:** Only 10.2% of reviewers consider alphabetical file ordering optimal, while the majority view it as suboptimal or neutral, revealing a consistent preference against the default ordering across users of different tools.

To further examine how such perceptions translate into practical confidence, we asked reviewers whether the order of files in code review tools had ever affected their ability to detect software bugs or issues. Among the 537 reviewers who followed alphabetical ordering, a majority expressed confidence in their bug detection regardless of file order (158 participants responded *"No"*). However, substantial uncertainty was evident, with 62.8% of respondents (42 *"Yes"* and 337 *"Maybe"*) indicating that the order of file presentations may affect their ability to detect bugs. This distribution is statistically significant ($\chi^2(2) = 382.0$, $p < 0.001$), highlighting a widespread lack of reviewer confidence in the effectiveness of alphabetical file ordering during code review. Among the reviewers who relied on alternative navigation strategies, 41.0% responded *"No,"* while 8.4% answered *"Yes"* and 50.6% selected *"May be."* These distributions were statistically significant ($\chi^2(2) = 382.0$ and 211.1, $p < .001$), revealing that skepticism toward alphabetical ordering persists even among those who do not use it, with nearly two-thirds (63.9%) expressing concern or uncertainty about its impact on bug detection (Cramér's $V = 0.28$, moderate effect). Given the widespread uncertainty about the effectiveness of alphabetical ordering, we next investigated the factors that influence reviewers' perceptions and preferences regarding file-ordering strategies.

> **Finding 2.2:** Over 63% of reviewers expressed uncertainty or concern about missing bugs, suggesting that the impact of file ordering on review effectiveness is both prevalent and substantively meaningful across reviewers.

*4.3.2* $\mathbf{RQ_{2.2}}$ *- Factors Influencing Perceptions.* To identify the factors shaping reviewers' perceptions toward file ordering, we performed regression analyses on three related outcomes (as shown in Table 1): i) optimality of alphabetical ordering, ii) desire for customizable ordering, and iii) perceived review-quality improvement under preferred ordering, while controlling for development and review experience, review frequency, prior experiences of missed bugs, and preference in alternative orderings.

**Perceived optimality of Alphabetical order.** The logistic regression model predicting whether reviewers considered alphabetical file ordering acceptable (*Yes/Neutral* = 1, *No* = 0) was statistically significant ($\chi^2(5) = 15.42$, $p < .001$), though it accounted for modest variance (Pseudo $R^2 = 0.01$). Only one predictor was significant: reviewers who believed alternative orderings improve effectiveness were less likely to view alphabetical orderings as acceptable (OR = 0.76, 95% CI [0.63, 0.91], $p = 0.003$). This indicates that each unit increase in agreement with the usefulness of non-alphabetical orderings corresponds to roughly 24% decrease in the likelihood of perceiving alphabetical ordering as acceptable. In contrast, demographic or experiential variables had no effect, indicating that the acceptance of alphabetical order reflects cognitive evaluations of efficiency rather than habitual or contextual familiarity.

**Table 1: Regression Models of Factors Influencing Reviewers' Perceptions of File Ordering**

| Outcome Variable | $\chi^2$(df) | $p$-value | Pseudo $R^2$ | Significant Predictors ($\beta$, OR, $p$) |
|---|---|---|---|---|
| Perceived optimality of alphabetical order | 15.42 (5) | <0.001 | 0.01 | Belief in usefulness of non-alphabetical ordering (−0.28, 0.76, <.05) |
| Desire for customizable ordering | 274.06 (7) | <0.001 | 0.16 | Missed bugs (0.47, 1.61, <.05); Expected higher feedback quality (0.61, 1.84, <.001); Belief in non-alphabetical benefit (1.01, 2.74, <.001) |
| Expected impact on review quality | 410.69 (5) | <0.001 | 0.19 | Missed bugs (1.00, 2.72, <.001); Belief in non-alphabetical benefit (1.26, 3.53, <.001); Desire for customization (0.61, 1.83, <.001) |

**Desire for customizable ordering.** We employed an ordinal logistic regression to examine reviewers' likelihood of desiring customizable file-ordering (*0 = No, 1 = Neutral, 2 = Yes*). The model was statistically significant ($\chi^2(7) = 274.06$, $p < .001$) and explained substantial variance (Pseudo $R^2 = 0.16$). As shown in Table 1, reviewers who had previously missed bugs expected higher feedback quality, or believed that non-alphabetical orderings improve outcomes, were significantly more likely to favor customization. Predicted probabilities indicate that approximately 66% of reviewers favored customization, 29% were neutral, and only 5% opposed it, suggesting a strong demand for tool-level flexibility in how changed files are presented during review. Reviewers who experienced inefficiencies or recognized cognitive benefits from context-driven sequencing are notably more inclined to support customizable ordering, independent of demographic or experiential factors.

**Table 2: Thematic Analysis of Reviewers' Rationales for Perceiving Alphabetical File Ordering as Suboptimal or Acceptable**

| Perception | Theme | Representative Examples |
|---|---|---|
| Suboptimal | Lack of semantic coherence (n=132) | *"Alphabetical order has no correlation with what changed or how.", "I'd rather start from the main logic and then move to the tests, not by file name.", ".... jump to other files to see what another component does, then come back, or search to see where a function is used in other files."* |
| | Cognitive fragmentation & context switching (n=133) | *"I often jump between files because one references a method in another.", "It's easy to lose the thread of logic when related files are far apart."* |
| | Review fatigue & uneven attention (n=101) | *"Files starting with later letters get less attention.", "By the time I reach the bottom, I'm tired; important files are easy to miss.", "If a more complex part of the code change is at the end of the order, your brain is already fatigued and can't focus as much as in the start of the review."* |
| | Misalignment with reviewer intent (n=214) | *"I always start with the core logic to understand what the change is about, then move to helper and test files. Alphabetical order makes that impossible.", "The main file that drives the change might be halfway down the list, buried under config or test directories.", "Alphabetical order forces me to break my flow. I can't just follow how the change propagates."* |
| Supportive/Neutral | Predictability & cognitive familiarity (n=293) | *"My IDE lists files alphabetically, so it feels natural to navigate reviews this way.", "Even if it's not meaningful, it's predictable and consistent across tools.", "When handling many files, alphabetical order makes it easy to navigate back and forth.", "It's not optimal, but at least it's predictable, so I know where to jump to."* |
| | Tolerable for small & routine reviews (n=203) | *"For small PRs, the order doesn't matter; I'll review all files anyway.", "As long as the change set is focused, alphabetical is fine, it keeps things simple."* |

**Expected impact of preferred ordering on review quality.** Finally, the third model employed an ordinal logistic regression to examine whether reviewers believed they would provide more thorough or constructive feedback if files were presented in their preferred order (*0 = No, 1 = Maybe, 2 = Yes*), showing statistically significant result ($\chi^2(5) = 410.69$, $p < .001$) and substantial variance (Pseudo $R^2 = 0.19$). As summarized in Table 1, reviewers who had previously missed bugs, believed that alternative orderings improve effectiveness, or desired customizable file sequencing were significantly more likely to expect higher-quality feedback when files were presented in their preferred order. Predicted probabilities based on the mean covariates indicate that approximately 19% of reviewers anticipated definite improvement, 59% were open to the possibility, and 22% did not foresee better outcomes. These results demonstrate that expectations for enhanced review quality are shaped by prior inefficiencies and cognitive evaluations of sequencing. Those who have encountered limitations with alphabetical ordering or perceive attentional and comprehension benefits from reordering are significantly more likely to anticipate providing higher-quality feedback when files are presented according to their preferred sequence.

> **Finding 2.3:** Reviewers consistently linked non-alphabetical and customizable ordering with improved comprehension and review effectiveness, suggesting that dissatisfaction with default ordering is cognitive rather than experience-driven.

*4.3.3* **RQ$_{2.3}$**- *Rationales Behind Perceptions.* Building on the reviewers' perceptions of the default alphabetical file ordering in **RQ$_{2.1}$**, we further explored why some reviewers considered alphabetical ordering suboptimal, while others viewed it as acceptable or remained neutral. Two individuals conducted a thematic analysis of 1,076 responses to identify recurring patterns and explanations. Table 2 illustrates the six identified themes (Cohen's $\kappa = 0.98$) along with representative participant examples. Among those who viewed alphabetical ordering as suboptimal (53.9%), four themes emerged, suggesting that dissatisfaction with alphabetical ordering arises primarily from cognitive rather than experiential factors. Reviewers considered alphabetical sequencing inconsistent with program logic, mentally demanding, and disruptive to their normal review process. Conversely, reviewers expressing supportive (10.2%) or neutral views (35.9%) emphasized predictability, familiarity, and pragmatic adequacy. They valued default ordering for its spatial stability across tools and its efficiency for small or routine reviews. This view reflects reviewers' tendency to minimize effort when the review scope is limited, prioritizing completion over the perfect order. However, several noted that tolerance declines as complexity increases, suggesting that alphabetical order is sufficient only under low cognitive load.

> **Finding 2.4:** Reviewers largely viewed alphabetical file ordering as misaligned with the cognitive processes underlying effective code review. It impeded logical comprehension and sustained attention, making it less suitable for large or cognitively demanding reviews. While its predictability offered convenience for smaller PRs, its lack of contextual meaning reduced its usefulness for understanding complex changes.

**Answer for RQ$_2$**– Reviewers widely regard alphabetical file ordering as cognitively inefficient and misaligned with how they reason about code. Their dissatisfaction is consistent across tools, driven more by cognitive evaluation than by experience, and reflects concerns about review effectiveness. Overall, developers favor context-driven or customizable file sequencing to better support comprehension, attention, and review quality.

## 4.4 RQ$_3$- File Ordering & Review Effectiveness

Building on the finding that reviewers perceive alphabetical ordering as suboptimal (**RQ$_2$**), we investigate how non-preferred file ordering influences review quality and effectiveness in practice.

Among reviewers who had previously expressed uncertainty or concern about missing defects (as noted in *Finding 2.2*), 46.6% explicitly acknowledged broader review quality issues related to change comprehension, feedback accuracy, and review completeness. Several participants described how arbitrary ordering disrupted their understanding of change dependencies, leading to premature or misinformed feedback. For instance, one reviewer noted: *"Yes, because it can take more time to understand the purpose of a file that is dependent on or called by another file further down the line. It can also lead to remarks being made at first that turn out to be moot points later in the code review."* Such fragmented comprehension often causes reviewers to comment based on incomplete information, only to realize that subsequent files make earlier observations invalid. As another participant explained, *"Sometimes I write a comment and later get to another file where the change makes my comment irrelevant."* This pattern results in low-precision feedback that wastes effort for both reviewers and PR authors while undermining the reliability of review outcomes.

In addition, reviewers emphasized that alphabetical ordering separates semantically related files, such as implementation and test code, thereby impeding verification of test adequacy. When test files appear near the end of the review, participants reported difficulties ensuring that functional changes were properly exercised. As one reviewer described, *"When reviewing very large changes, correlating logic changes to test changes is very burdensome as they are usually in different parts of the patch order."* Similarly, the intermixing of trivial or auto-generated files with critical code modifications forced reviewers to navigate through irrelevant content before reaching meaningful changes. This problem not only disrupts cognitive flow but also leads to frustration and wasted effort. As one reviewer admitted, *"I don't read every single line because I want to get a glimpse of the important parts first. If I keep searching for the important changes and they turn out to be at the end, I have to process many unrelated files, which destroys the flow and wastes a lot of time."*

**Finding 3.1:** Non-preferred file ordering disrupts review context, leading to premature or irrelevant feedback and obscuring critical changes and test adequacy.

Finally, we examined how reviewers cope with non-preferred alphabetical file ordering by mentally reordering files or using external tools. A majority (over 59%) reported not reordering, often citing limited tool support or minimal perceived benefit. However, about 41% reported mental reordering or using external aids such as IDEs, scripts, or browser extensions. This behavior suggests that reordering reflects a fundamental cognitive adaptation rather than an optional enhancement. Several reviewers explicitly described constructing temporary mental sequences to maintain context: *"I reorder files in my head constantly. As I build up context, I scan files for dependencies and inconsistencies. Sometimes a file seen earlier makes more sense after reading one further down."* Others acknowledged cognitive limitations, highlighting the working-memory constraints involved in mental reordering: *"I usually cache the next 3–6 elements of my own order in my mind."* When web-based interfaces became cognitively or navigationally overwhelming, reviewers frequently checked out code locally to regain control over navigation and semantic grouping. As one respondent explained: *"I often clone the branch in my editor so I can use LSP to trace references and cross-reference against diffs in GitLab. It is slow and tedious, but it helps maintain context."* Yet, this workaround introduces trade-offs. A reviewer noted: *"Using my local IDE works, but it does not show recent changes against the base branch as nicely as GitHub's UI. Pulling locally also makes it harder to identify what actually changed."* Even within web interfaces, participants actively attempted to manage order using collapse, mark-as-viewed, or file-type filtering features, although these were viewed as limited solutions. For instance, one reviewer described: *"Our private tool allows us to filter files by type or extension so I can hide unimportant parts such as tests, styles, or documentation and focus on the core logic."* Collectively, these findings reveal that reviewers spend considerable effort reconstructing a logical order to preserve comprehension, manage cognitive load, and maintain efficiency. Manual and tool-mediated reordering, therefore, represents an essential compensatory behavior that exposes a deep misalignment between current review interfaces and the way developers cognitively organize change comprehension.

**Finding 3.2:** Reviewers often mentally or manually reorder files to restore logical flow. Yet, these efforts are cognitively demanding and inefficient, revealing a clear misalignment between tool ordering and reviewer reasoning.

**Answer for $\mathbf{RQ_3}$** – File ordering influences review outcomes by shaping how reviewers comprehend and navigate changes, with suboptimal ordering increasing effort and attention loss.

## 4.5 $\mathbf{RQ_4}$- Multi-file PRs Challenges

Having shown that file ordering affects review effectiveness (**$\mathbf{RQ_3}$**), we now examine the broader challenges reviewers face in multi-file PRs, distinguishing between issues inherent to multi-file reviews from those made worse by poor file ordering.

*4.5.1* $\mathbf{RQ_{4.1}}$ *- General Multi-File Review Challenges.* To investigate common Challenges (Cs) in multi-file PR reviews, we analyzed 2,093 responses from 1,327 reviewers, categorizing them into six themes. Two evaluators independently validated the classifications with high agreement (Cohen's $\kappa$ = 0.88). Nearly all participants reported at least one challenge; only nine reported no significant difficulties, citing factors such as familiarity with the codebase, review experience, or effective strategies.

**C1: Understanding change impact across files (836 responses):** 37.7% of reviewers reported difficulty assessing the overall impact of changes in multi-file PRs. Challenges include understanding developer intent, inter-dependencies among modifications, and broader architectural effects, especially when critical context lies outside the modified files or spans unrelated components. One reviewer described the challenge as *"understanding the intent of the changes/what the person wants to do,"* while another stressed *"understanding how exactly different parts interact with each other based on all the possible use-cases."* Reviewers also noted the cognitive burden of synthesizing fragmented modifications when implicit dependencies remain hidden, as reflected in *"Getting the whole changeset into my head at once."*

**C2: Tracking interfile dependencies (439 responses):** 28.5% of reviewers highlighted the difficulty in maintaining continuity across the review due to inter-file dependencies. This challenge is pronounced when changes span multiple files, requiring constant mental reconstruction to understand cross-file effects and maintain context. Reviewers reported difficulties tracing how changes in one file affect others, identifying relationships, and remembering what has been reviewed. One noted the difficulty of *"keeping track of what I have reviewed when I am not reviewing each file linearly,"* while another pointed to *"jumping between files/changes, e.g., to check whether some change in one file matches a change in another file."* Others mentioned re-reviewing content due to lost context or the inability to locate call sites or definitions, underscoring the friction and cognitive load in navigating large PRs.

**C3: Prioritizing attention across files (436 responses):** 30.9% of reviewers reported challenges in allocating attention effectively across files, especially when identifying those containing critical and core logic changes. This often led to disproportionate focus on minor edits, such as formatting or import order, while critical changes remained under-reviewed. Participants expressed the need for more effective methods to identify essential files and align the review focus with their expertise. One noted the challenge of *"figuring out which file is the 'main' one driving the changes vs which are dependencies being updated,"* while another described the difficulty of sustaining focus: *"Often you start reviewing it thoroughly, but by the end you can run out of energy."* Excessive non-essential changes also contributed to cognitive overload, reducing reviewers' ability to ensure adequate attention to high-impact code.

**C4: Ensuring completeness of changes (312 responses):** This challenge, cited by 21.8% of reviewers, involves verifying that all necessary files have been updated and that no critical modifications have been missed. Reviewers often struggle to detect both intentional and unintentional omissions, such as missing logic, unmodified dependencies, absent test cases, or overlooked related files. Participants expressed concerns like *"there are missing files that should have been in the PR"* and *"a file not included in the PR should have also been modified."* These insights reveal the difficulty of identifying silent gaps in the changeset, especially without full test coverage or visibility into cross-file dependencies. One reviewer captured this issue as catching *"errors of omission: find which files/-lines (if any) should have also changed but were not modified."*

**C5: Tooling and UI limitations (37 responses):** Reported by 2.8% of reviewers, this challenge reflects shortcomings of code review tools that impede efficiency and thoroughness. Reviewers cited issues such as delayed responsiveness, limited context display, and poor navigation in large diffs. GitHub's interface was often criticized for collapsing large files, rendering slowly, and hiding crucial context. One participant described the experience as *"Fighting the terrible GitHub UI,"* while another noted, *"It usually shows very little context around the lines that changed. I spend quite some time expanding the lines of code it shows in each file."* Such constraints disrupt workflow and often require reviewers to adopt alternative platforms or workarounds to sustain review effectiveness.

**C6: Poorly scoped or structured PRs (24 responses):** Issues related to PR scope and organization were identified by 1.8% of reviewers. These challenges arise when changesets are overly large, mix unrelated modifications, or lack clear organization. Such practices violate fundamental engineering principles, increase reviewer fatigue, and diminish review effectiveness. One participant noted frustration with *"bloated PRs with non-atomic changes and bloated classes/methods with single responsibility principle violation,"* while another stated, *"PRs not being small enough in scope should be split into multiple PRs."* These issues obscure change rationale and hinder the delivery of focused feedback, often prompting reviewers to request refactoring before proceeding.

> **Finding 4.1:** Reviewers face multifaceted challenges when reviewing multi-file PRs, including understanding cross-file impact, tracking dependencies, prioritizing attention, and ensuring completeness, issues often compounded by tool limitations and poorly scoped or structured PRs.

*4.5.2* **RQ$_{4.2}$** - *Ordering-Specific Difficulties.* Besides the multi-file PR review challenges, we collected insights about reviewer Difficulties (Ds) when PR files lack their preferred ordering (Figure 1). We received 2,231 responses, which were manually categorized into five groups (Cohen's $\kappa$ = 0.83) and discussed as follows.

**D1: Time and navigation inefficiency (695 responses):** Sub-optimal file ordering leads to significant time inefficiencies and navigation burdens, resulting in workflow disruptions and compromised review quality, as reviewers often address minor issues before encountering critical ones. Many report spending excessive effort *"finding the more critical (security, performance, etc.) file"* and even physical discomfort such as *"finger pain for all the scrolling."* Technical constraints worsen the experience - users with low bandwidth face delays downloading irrelevant changes, while large PRs strain browsers and hinder visibility. To cope, several reviewers re-order columns or use external tools like *'TortoiseGit'*, underscoring the inadequacy of current file presentation.

**D2: Cognitive load and mental strain (499 responses):** Non-preferred file organization imposes significant cognitive overhead, forcing reviewers to invest additional mental effort to comprehend changes. This leads to confusion regarding baseline comparisons and modification rationale, inefficient review patterns, repeated file visits, and invalidated assumptions, *"I make some assumption about how the code works early, that gets invalidated, and I have to restart my review from that point."* This cognitive load intensifies when reviewers must actively locate critical changes rather than encountering them sequentially.

**D3: Contextual understanding and change relationships (473 responses):** When file order lacks contextual relevance, reviewers struggle to comprehend change scope and logical connections across files, often compensating by manually *"scrolling through all the files to get the bigger picture first."* In such cases, reviewers often rely on commit order, *"the ordering of the commits tell the changes,"* emphasizing the importance of meaningful sequencing. Fragmented presentation further impairs understanding of implementation and test relationships, making it harder to *"relate modifications to created tests"* or assess test results when relevant changes are not co-located.

**D4: Review quality and thoroughness risks (369 responses):** Poor organization of files threatens code review quality and completeness, potentially undermining code integrity assurance. Reviewers acknowledge the risk of providing *"sub-quality review"* when frustrated by disorganized presentation, with some admitting they *"could definitely grow impatient and start overlooking until it gets to the point"* where critical issues are missed. This leads to problematic effort misallocation, as reviewers spend disproportionate time *"over-reviewing less important files"* while rushing through critical changes. Fatigue-induced shortcuts further compromise review completeness, with reviewers stating *"I just skip if I get too tired."* Such exhaustion can also trigger avoidance behaviors, with some reviewers *"putting the review off"* entirely.

**Table 3: Expected File Ordering in PRs Mentioned by Reviewers**

| Preferred Ordering | Description | Count |
|---|---|---|
| File sorting by dependency tree | Ordering files by their logical or execution dependencies, following the dependency tree, call chain, or control flow from higher-level modules/interfaces down to low-level implementations and tests (e.g., interface → implementation → tests → downstream changes). | 489 |
| File clustering by module | Grouping files based on their semantic or module relationships by file type, feature, or purpose (e.g., production code, tests, documentation), including similarity of change and syntax-aware semantic similarity. | 452 |
| Hyperlinked related files | Enhancing navigability within PRs with features that facilitate seamless transitions between related or dependent files (both modified and unmodified), including hyperlinked navigation to definitions, uses, and contextual code. | 427 |
| Priority by change criticality | Prioritizing files by their importance, risk, or impact on the pull request's goals, with critical changes and core system components appearing first, while trivial or mechanical changes are prioritized. | 405 |
| Change size ordering | Ordering files by the magnitude of changes (lines of code modified), with some preferring smaller changes first for quick processing and others preferring larger changes first to tackle substantial work upfront. | 20 |
| Author-specified ordering | Allowing authors to control file ordering in PRs, leveraging their deep understanding of the changes to present files in the most logical and review-friendly sequence, often with explanations in the PR description to guide reviewers. | 19 |
| Commit/time-based ordering | Following the author's development chronology by reviewing files in commit order or modification time allows reviewers to understand the logical progression of changes and maintain the same mindset as the developer during implementation. | 8 |
| AI/tool-assisted ordering | Using AI tools to intelligently analyze and prioritize code changes, providing summaries of critical modifications and contextual relationships between files to surface the most important aspects of a PR | 3 |

**D5: Emotional impact and motivation loss (169 responses):** Suboptimal file representations also create emotional barriers that undermine reviewer engagement and motivation, fostering avoidance behaviors. This emotional strain appears as *"procrastination,"* with reviewers delaying or avoiding reviews when confronted with overwhelming or confusing file structures, with some stating they *"just skip if I get too tired of just looking."* This psychological burden transforms constructive technical processes into sources of stress and avoidance, representing a critical yet overlooked aspect of review effectiveness. The emotional impact extends beyond individual reviews, potentially creating lasting negative associations that affect long-term team collaboration and code quality maintenance.

We also found 26 responses showing complete indifference to file ordering, with reviewers stating no preference or considering it unimportant. These participants emphasized individual changes over file-level organization during the review.

> **Finding 4.2:** Poor file ordering in PRs significantly impairs review quality and effectiveness, leading to time inefficiencies, increased cognitive strain, and even avoidance behaviors.

Figure 2 compares the challenges reviewers face in multi-file PRs with their preferred file ordering strategies, both analyzed by years of code review experience. On the left, it shows that reviewers with 8–11 years of experience consistently reported the highest number of challenges, suggesting heightened sensitivity to inefficiencies. Those with 16 or more years also reported frequent difficulties, indicating persistent obstacles even among experienced reviewers. In contrast, reviewers with 0–3 years of experience reported challenges less often, likely due to limited exposure or involvement in complex reviews. Review difficulties, such as understanding change impact, cognitive load, and navigation inefficiency, were common across all groups. However, they were especially prominent among mid to senior reviewers (4–15+ years). Less experienced participants reported no significant issues more frequently, possibly reflecting their limited engagement with large PRs.

**Answer for $\mathbf{RQ_4}$–** Reviewers experience several challenges in multi-file PRs, such as understanding cross-file relationships, tracking dependencies, prioritizing key changes, and verifying completeness. These issues are often made harder by limited tool support and large, unfocused PRs. Poor file ordering further amplifies these problems, increasing review time, cognitive effort, and fatigue, and reducing overall review quality.

## 4.6 $RQ_5$- Reviewer Expectations for Ordering

Building on the cognitive and usability challenges identified in $\mathbf{RQ_4}$ and the perceptual issues revealed in $\mathbf{RQ_2}$-$\mathbf{RQ_3}$, we now examine what improvements reviewers desire for multi-file PR review tools.

Table 3 summarizes reviewers' preferred file-ordering strategies (Cohen's $\kappa$ = 0.91), while the right side of Figure 2 illustrates how these preferences vary by experience. File sorting by dependency and clustering by module are the most preferred strategies across all groups, with peak support from reviewers with 8–11 years of experience, consistent with their emphasis on structured review practices. Reviewers with 4–15 years of experience favored logical, context-aware orderings, including prioritization by change criticality. Preferences among the most experienced reviewers (16+ years) remained strong, however, slightly declined, suggesting greater flexibility. Reviewers with 0–3 years also showed a preference for structured orderings, although at lower rates. Less common strategies, such as author-specified, change size, directory-based, and commit/time-based orderings, were selected by a few participants, with no strong alignment to any specific experience level.

Finally, most reviewers (63.17%) supported customizable file ordering, while only 7.78% opposed it, citing no tangible benefits. Support was strongest among reviewers with 8–11 and 16+ years of experience (over 46%), who emphasized that flexibility would help organize complex reviews and concentrate effort on high-impact changes. Many anticipated modest yet meaningful improvements, with an estimated 10–25% increase in review speed and fewer overlooked issues, achieved through better prioritization rather than shorter review times. As one participant noted, *"It may not save much time, but I believe it would result in a lower bug rate,"* while another added, *"Probably spend the same amount of time on the review but spend more time on the important parts."* These findings extend the earlier behavioral evidence (**RQ$_{2.2}$**) by showing that experienced reviewers translate their positive perception of non-alphabetical orderings into a clear expectation for practical tool support enabling adaptive, context-aware file organization.

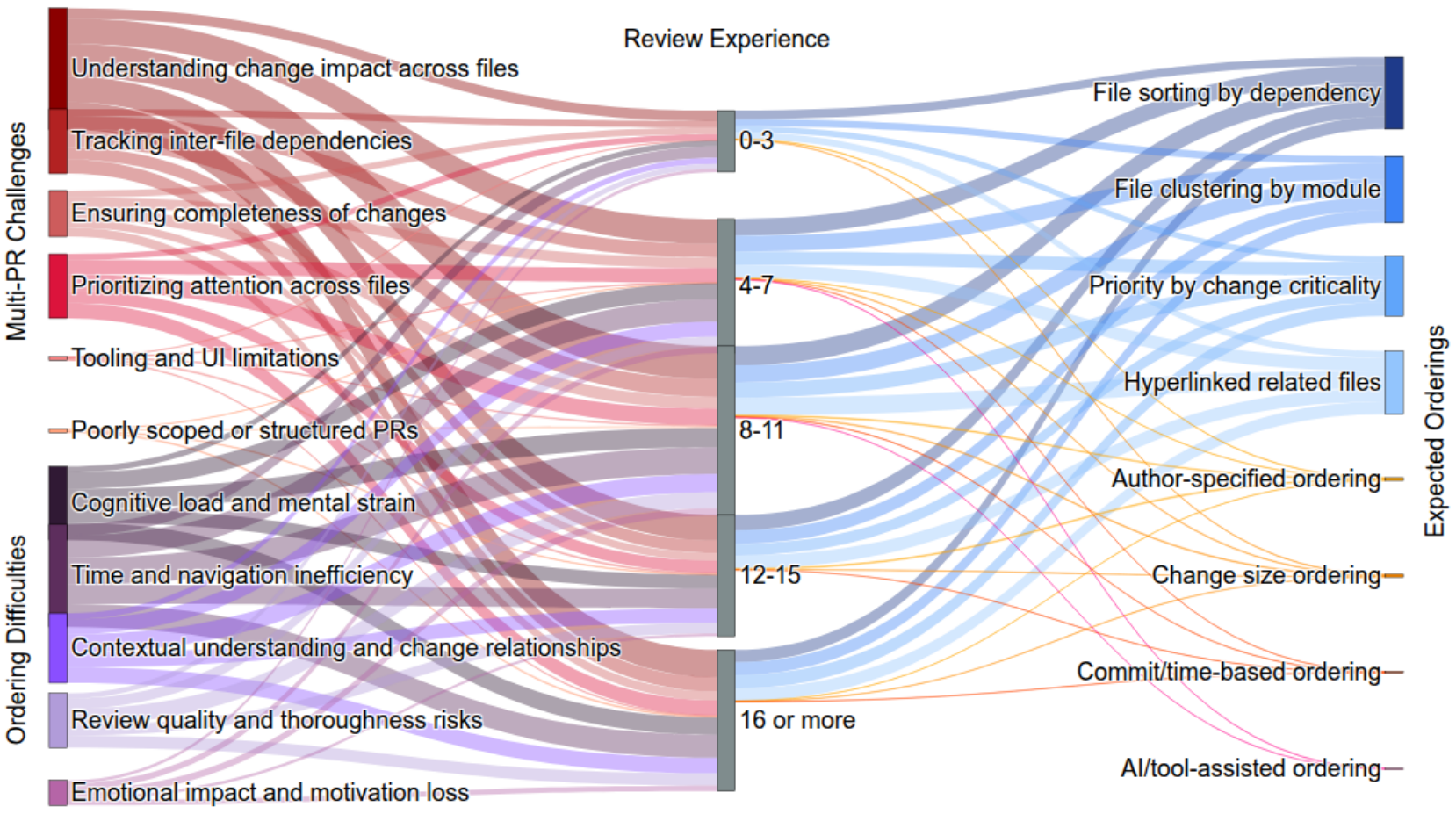

**Figure 2: Challenges in Reviewing Multi-File PRs and File Ordering Expectations Across Experience Levels**

**Answer for RQ$_5$–** Reviewers favored structured and customizable file ordering that reflects dependencies or logical relationships among files. Experienced reviewers particularly valued this flexibility for managing complex reviews and focusing on critical changes, emphasizing that such customization enhances comprehension and review precision rather than speed.

# 5 Discussion

## 5.1 Tool-Cognition Misalignment

This study reveals a fundamental misalignment between tool-imposed file ordering and reviewers' cognitive strategies for reasoning about change. While nearly all major review platforms default to alphabetical sequencing, most reviewers reject it as optimal. This persistence reflects a design failure rather than a user preference, i.e., reviewers use alphabetical ordering not by choice but by constraint. Such ordering adds extraneous cognitive load [10, 41, 42, 94], forcing reviewers to mentally reconstruct semantic relationships rather than focusing on comprehension. Prior studies have shown that presentation format directly affects code comprehension and cognitive load [34, 40, 71]. Although reviewers attempt mental or manual reordering, working memory limits [25, 26] make such coping strategies fragile, especially in large PRs. Reviewers consistently noted in our survey how alphabetical order disrupts hierarchical comprehension [90], often separating related files (e.g., implementation–test pairs) and inducing misaligned mental models [74]. In code review, this leads to premature judgments and lower-quality feedback driven by tool-imposed cognitive constraints, while well-structured, semantically grouped presentations yield higher performance [36, 55]. In addition, reviewers' concerns about missing bugs align with research on attention decrement in sequential tasks, sustained attention deteriorates across extended sequences [38, 66]. Our results show that cognitive fragmentation-context rebuilding across unrelated files accelerates this decline. Hierarchical navigation and perceptual grouping mitigate such load [35, 92]. Moreover, alphabetical persistence reflects path dependence [4] and legacy affordances from early lexicographic file systems [68], privileging computational over cognitive efficiency [28, 93]. Human-factors research links usability alignment to productivity and satisfaction [21, 79], while misaligned CASE tools are often abandoned [89]. The avoidance, fatigue, and frustration reported here mirror threats to the social sustainability of code review practices [30]. These findings highlight the need for reviewer-centric tools that align file presentation with cognitive workflows, using dependency-aware grouping, semantic clustering, and customizable ordering to preserve context, reduce attention loss, and enhance review effectiveness.

## 5.2 Reviewer-Centric Tool Design

Our findings call for a shift from tool-centric to reviewer-centric interface design in code review. Current platforms optimize for implementation simplicity as alphabetical ordering requires no dependency analysis and scales easily. Yet this efficiency overlooks the cognitive processes that determine review effectiveness. **RQ$_5$** (Section 4.6) shows a clear consensus for desired improvements. Reviewers consistently favor dependency-aware, semantically grouped, and customizable file ordering that reflects how they build mental models of code changes. These preferences are empirically grounded rather than speculative. The strong support for customization indicates that no single ordering is suitable for all contexts or review goals, suggesting that tools should provide adaptive interfaces rather than rigid defaults. The preferences expressed align closely with established cognitive research on hierarchical mental model construction [90], working memory limitations [25, 26], and the importance of perceptual grouping in information processing [92]. These directions are technically feasible, as modern static analysis and code intelligence can already extract dependencies, cluster related files, and compute change-impact metrics [3, 39, 58]. Moreover, reviewer-centric design aligns with broader movements in human-centered software engineering [56, 59, 77] and cognitive-driven development [72]. These approaches emphasize aligning tools with developers' cognitive and emotional processes, thereby reducing friction and improving reasoning flow. Recent research on developer experience (DevEx) and productivity further shows that minimizing cognitive load, shortening feedback loops, and supporting mental model continuity directly enhance performance and satisfaction [21, 67, 76]. Advancing this vision requires collaboration among researchers, tool builders, and practitioners to translate empirically validated preferences into evidence-based interfaces that enhance comprehension, efficiency, and reviewer well-being.

## 5.3 Implications

This study offers implications that extend across organizational practice, tool design, and future research. **For development teams,** the findings underscore the importance of disciplined PR authoring. Atomic changes, logical file grouping, and clear descriptions can mitigate the cognitive load imposed by suboptimal ordering. Teams should also manage review effort strategically by limiting PR size, rotating responsibilities to prevent reviewer fatigue, and training reviewers to recognize positional bias and attention decrement. **For organizations and tool adopters,** the strong demand for customizable ordering indicates an opportunity for improvement. Engaging with platform maintainers, contributing to open-source review ecosystems, or developing internal extensions can translate reviewer preferences into tangible benefits. Cognitive-aligned review environments may yield measurable improvements in both review quality and developer satisfaction. **For researchers and tool builders,** $RQ_5$ highlights a promising space for adaptive and intelligent review interfaces. Future work should evaluate machine-learned ordering models, examine cognitive-load dynamics through physiological and eye-tracking methods, and study longitudinal effects on review accuracy and engagement. Future studies should combine controlled and real-world evaluations to assess the effectiveness of cognitively aligned designs [37, 54, 57].

Beyond code review, these insights contribute to a broader understanding of cognitive ergonomics in software engineering [31, 78]. Tool design actively shapes reasoning, collaboration, and satisfaction, reinforcing the need for human-centered approaches in software engineering research, practice, and education.

# 6 Threats to Validity

**Construct Validity.** The main construct validity threat relates to how we operationalized and measured the concepts in our survey. The notions of *"file ordering optimality," "review effectiveness,"* and *"cognitive load"* are subjective and maybe interpreted differently by participants. To mitigate this threat, we carefully designed survey questions in accordance with established guidelines for empirical software engineering research [53, 73]. We conducted a two-phase expert review with industry practitioners and researchers to ensure questions accurately captured the intended constructs (as discussed in Section 3.2.2). Additionally, we used both closed-ended questions with clear scale definitions and open-ended questions to allow participants to express their interpretations, enabling triangulation of responses and validation of our construct measurements.

**Internal Validity.** Our study relies on self-reported perceptions and behaviors from survey participants, introducing potential biases. *Recall bias* may affect participants' ability to accurately recall specific review experiences or how often they encounter certain challenges [43, 83]. *Social desirability* bias could also lead participants to report practices they perceive as *"correct"* rather than their actual behaviors [44, 56]. To mitigate these risks, we designed questions to focus on general patterns over isolated incidents and assured participants of complete anonymity to encourage honest responses. We also applied methodological triangulation by comparing responses across related questions and aligning closed-ended items with corresponding open-ended explanations [63]. *Selection bias* represents another internal validity threat. Our recruitment method (emailing publicly available addresses from GitHub projects and snowball sampling) may have attracted developers who are more engaged with open-source practices or with stronger opinions about code review tools, as documented in GitHub-based surveys [8, 51]. Developers with publicly accessible email addresses might differ systematically from the broader population of code reviewers. However, we achieved a substantially diverse sample, with 1,355 responses from 67 countries and participants ranging from 0-3 to 16+ years of review experience, which helps mitigate this concern. Finally, our survey's cross-sectional design captures perceptions at a single point in time and may not reflect evolving practices or the effects of emerging tool improvements [29]. Furthermore, while statistical analyses reveal associations between experience levels, ordering preferences, and review perceptions, they cannot establish causal relationships [95]. Our results should therefore be interpreted as correlational rather than causal evidence.

**External Validity.** The generalizability of our findings maybe limited by the sampling frame and participant demographics. We recruited contributors from 182 popular open-source projects hosted on Apache and GitHub. While this provided a diverse sample within the open-source ecosystem, our findings may not generalize to developers working exclusively in closed-source or proprietary environments where review practices, tools, and constraints differ substantially. Moreover, our survey assumes familiarity with modern code review tools that default to alphabetical file ordering (e.g., GitHub, GitLab). Results may not apply to teams using specialized review tools with different default orderings or organizations with established review protocols that override tool defaults. Additionally, cultural and organizational factors influencing review practices may vary beyond what our survey captured.

# 7 Conclusion

This study advances the understanding of how review-tool design influences developers' cognitive performance and review efficiency. Drawing on a large-scale survey of 1,355 professionals across 182 popular open-source projects, we found widespread dissatisfaction with alphabetical file ordering, largely due to difficulties in tracking inter-file dependencies and prioritizing critical changes, alongside a strong preference for semantically or logically grouped sequences. Respondents further reported that alphabetical order fragments context, increases navigation effort, and undermines confidence in defect detection. These results suggest that attention alone is insufficient, as effective code review requires cognitively aligned workflows and supportive tools. Our findings underscore the need to rethink file presentation in review platforms and to design context-aware, reviewer-centric environments that better reflect developers' mental models. As future work, we plan to explore intelligent file-ordering heuristics and deeper integration of dependency analysis to support more effective and defect-reducing review processes.

# Acknowledgments

This research is supported in part by the Natural Sciences and Engineering Research Council of Canada (NSERC) Discovery Grants program and by the industry-stream NSERC CREATE in Software Analytics Research (SOAR). We also gratefully acknowledge the anonymous survey respondents for their valuable time and insights.